\documentclass[twocolumn,nofootinbib,prd,showpacs,floatfix]{revtex4}
\usepackage{epsfig}

\begin{document}
\title{First hints of large scale structures in the ultra-high energy sky?}
\author{A. Cuoco$^1$, G. Miele$^1$, and P.~D.~Serpico$^2$}

\affiliation{ $^1$Universit\`{a} ``Federico II", Dipartimento di
Scienze Fisiche, Napoli, Italy \& INFN Sezione di Napoli\\
$^2$Particle Astrophysics Center, Fermi National Accelerator
Laboratory, Batavia, IL 60510-0500 USA}

\begin{abstract}
The result of the recent publication \cite{Kachelriess:2005uf} of a
broad maximum around 25 degrees in the two-point autocorrelation
function of ultra-high energy cosmic ray arrival directions has been
intriguingly interpreted as the first imprint of the large scale
structures (LSS) of baryonic matter in the near universe. We analyze
this suggestion in light of the clustering properties expected from
the PSCz astronomical catalogue of LSS. The chance probability of
the signal is consistent within 2 $\sigma$ with the predictions
based on the catalogue. No evidence for a significant
cross-correlation of the observed events with known overdensities in
the LSS is found, which may be due to the role of the galactic and
extragalactic magnetic fields, and is however consistent with the
limited statistics. The larger statistics to be collected by the
Pierre Auger Observatory is needed to answer definitely the
question.
\end{abstract}

\pacs{98.70.Sa    
\hfill DSF-33/2006, FERMILAB-PUB-06-372-A}

\date{\today}
\maketitle
\section{Introduction}
The origin of ultra-high energy cosmic rays (UHECRs) is still an
open problem and, at the present, two different classes of models
compete to explain the most energetic  events observed. In
``bottom-up" mechanisms the acceleration up to extreme energy occurs
in suitable astrophysical environments, whereas in ``top-down"
scenarios UHECRs are produced by the decay or annihilation of
super-massive relic particles in the halo of our Galaxy or by
cosmological diffuse topological defects. The observation that UHECR
arrival directions (in particular at energies $E\agt 8\times
10^{19}\,$eV) may cluster according to the underlying large scale
structure (LSS) of the universe would represent a clear evidence in
favor of the ``bottom-up" mechanisms, and should co-exist with the
flux suppression known as the Greisen-Zatsepin-Kuzmin (GZK) effect
\cite{Greisen:1966jv,Zatsepin:1966jv}. The challenging and
fascinating problem of determining at which energy (if any)
astronomy with charged particles becomes possible is thus strictly
related to the identification of the sources of UHECRs, which in
turn would constrain the galactic and extragalactic magnetic fields
as well as the chemical composition of the primaries. The latter
point is an important prerequisite to use UHECR data to study
particle interactions at energy scales otherwise inaccessible to
laboratory experiments.

It is well known that the chances to perform cosmic rays astronomy
increase significantly at extremely high energy, in particular due
to the decreasing of deflections in the galactic/extragalactic
magnetic fields. Moreover, at $E\agt 4-5\times 10^{19}\,$ eV the
opacity of the interstellar space to protons drastically grows due
to the photo-pion production $p+\gamma_{\rm CMB}\to\pi^{0(+)}+p(n)$
on cosmic microwave background (CMB) photons (GZK effect). A similar
phenomenon at slightly different energies occurs for heavier
primaries via photo-disintegration energy losses. Above this range
of energy, most of the flux comes from sources within a distance of
few hundred Mpc (see e.g. \cite{Harari:2006uy}), and this should
facilitate the source identification. Thus, the GZK feature in the
spectrum and the large-scale anisotropy should be correlated
signatures. The drawback is that in the trans-GZK regime the flux is
greatly suppressed, even beyond the expected power-law extrapolation
of UHECR spectrum, and instruments with huge collecting areas are
required to accumulate sufficient statistics to attack the problem.
A final answer is expected when the Pierre Auger Observatory
\cite{Auger,Abraham:2004dt} will have detected enough events.

Until now, the experiments of the previous generation have collected
${\cal O}$(100) events above $E\agt 4-5\times 10^{19}\,$ eV, and one
may wonder if any useful hint of the UHECR sources already hides in
the available catalogues. In the recent publication
\cite{Kachelriess:2005uf}, the authors found some evidence of  a
broad maximum of the two-point autocorrelation function of UHECR
arrival directions around 25 degrees. The evidence was obtained
combining the data with energies above $4\times 10^{19}$ eV of
several UHECR experiments, after an {\it a priori} adjustment of
their energy scale. This signal becomes significant only when
several data-sets are added, but it is not caused solely by an
incorrect combination of the exposure of different experiments. Both
the signal itself and the exact value of the chance probability have
to be interpreted with care, since the authors did not fix a priori
the search and cut criteria. Although the nominal value of the
chance probability for the signal to arise from random fluctuations
is around $0.01\%$, when taking into account a penalty factor of 30
they estimated the `` true chance probability" of the signal to be
of the order of $P\sim 0.3\%$. The authors suggest that, given the
energy dependence of the signal and its angular scale, it might be
interpreted as a first signature of the large-scale structure of
UHECR sources and of intervening magnetic fields.

The aim of this work is to test their qualitative interpretation of
the result on the light of the signal expected if UHECR data reflect
the large scale structure distribution of galaxies in the nearby
universe. In Ref. \cite{Cuoco:2005yd}, we have performed a forecast
analysis for the Pierre Auger Observatory, to derive the minimum
statistics needed to test the hypothesis that UHECRs trace the
baryonic distribution in the universe. Assuming proton primaries, we
found that a few hundred events at $E\agt5\times 10^{19}\,$eV are
necessary at Auger to have reasonably high chances to identify the
signature, independently of the details on the injection spectrum.
In this work we calculate the expected signal in terms of the
autocorrelation function as in \cite{Kachelriess:2005uf} for the
presently available statistics, and discuss quantitatively how well
predictions based on the LSS distribution can reproduce their
findings. The method and the results obtained are presented in Sec.
\ref{sect2}; in Sec. \ref{conclusions} we briefly discuss our
findings and conclude.

\section{UHECR clustering on medium scales and LSS}\label{sect2}

In our analysis, we closely follow the approach reported in
\cite{Kachelriess:2005uf}, using a similar dataset extracted from
available publications or talks of the AGASA
\cite{Hayashida:2000zr}, Yakutsk \cite{yakutsk}, SUGAR
\cite{Winn:1986un}, and HiRes collaborations
\cite{Abbasi:2004ib,Hires2}. In particular, in order to match the
flux normalization of HiRes, the energies of the AGASA data must be
rescaled downwards by $\sim 30\%$, while the energies of Yakutsk and
SUGAR data by $\sim 50\%$. We address the reader to
\cite{Kachelriess:2005uf} for further details.

We define the
(cumulative) autocorrelation function $w$ as a function of the
separation angle $\delta$ as
\begin{equation}
w(\delta)=\sum_{i=2}^N\sum_{j=1}^{i-1}\Theta(\delta-\delta_{ij}),
\end{equation}
where $\Theta$ is the step function, $N$ the number of CRs
considered and $\delta_{ij} =\arccos(\cos\rho_i\cos\rho_j +
\sin\rho_i \sin\rho_j \cos(\phi_i -\phi_j))$ is the angular distance
between the two cosmic rays $i$ and $j$ with coordinates $(\rho,
\phi)$ on the sphere. We perform a large number $M\simeq 10^5$ of
Monte Carlo simulations of $N$ data sampled from an uniform
distribution on the sky and for each realization $j$ we calculate
the autocorrelation function $w_j^{\rm iso}(\delta)$. The sets of
random data match the number of data for the different experiments
passing the cuts after rescaling, and are spatially distributed
according to the exposures of the experiments.
 The formal
probability $P(\delta)$ to observe an equal or larger value of the
autocorrelation function by chance is
\begin{equation}
P(\delta)=\frac{1}{M}\sum_{j=1}^M\Theta[w_j^{\rm iso}(\delta)-w_\star(\delta)],
\end{equation}
where $w_\star(\delta)$ is the observed value for the cosmic ray
dataset and the convention $\Theta(0)=1$ is being used. Relatively
high values of $P$ {\it and} $1-P$ indicate that the data are
consistent with the null hypothesis being used to generate the
comparison samples, while low values of $P$ {\it or} $1-P$ indicate
that the model is inappropriate to explain the data. Note also that
by construction the values at different $\delta$ of the function
$P(\delta)$ are not independent. Nonetheless, studying the
cumulative distribution function (as opposed to the differential
one) is the only realistic way to extract information in a low
statistics ``noisy" sample. In addition, an autocorrelation
study---differently from the approach of Ref. \cite{Cuoco:2005yd}
where a $\chi^2$-analysis was used---only relies on the clustering
probability in the data, while any directionality in the signal is
lost. Although providing less compelling evidence, this method has
the advantage of being more robust towards large magnetic
deflections. As long as the energy and the charge of primaries from
the same source are similar, their relative displacement should be
small compared with the absolute displacement with respect to their
sources. Thus it is natural to expect that the first (although more
ambiguous) hints of a signal may come from the study of $w(\delta)$.

\begin{figure}[!tb]
\vspace{-0.9pc}
\begin{center}
\begin{tabular}{c}
\epsfig{figure=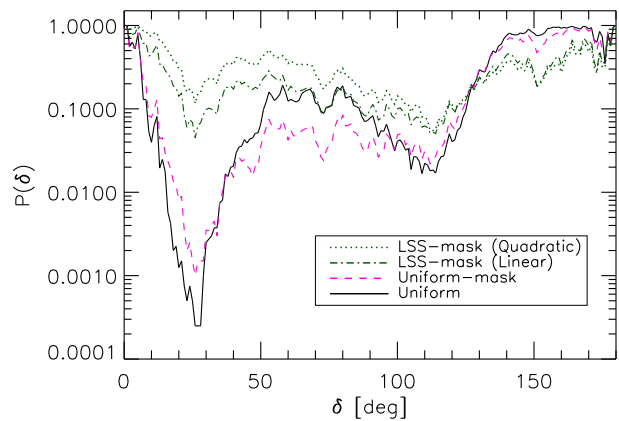,width=1.0\columnwidth}
\end{tabular}
\end{center}
\vspace{-0.9pc} \caption{The solid line shows the chance probability
$P(\delta)$ to observe an equal or larger value of the
autocorrelation function as function of the angular scale $\delta$
for the combination of experimental data of
Hires+AGASA+Yakutsk+SUGAR as described in the text. The dashed
purple line is the same signal, when cosmic rays falling in the PSCz
catalogue mask are disregarded. The dot-dashed green line is the
same quantity, if the random events are sampled according to the LSS
distribution, instead of an uniform one. The dotted green line is
the result for a sample proportional to the square of the LSS
distribution.} \label{w_delta_check}
\end{figure}

Figure \ref{w_delta_check} summarizes our main results. The solid,
black curve shows that under the same assumptions of Ref.
\cite{Kachelriess:2005uf}, we obtain the same behavior for the
function $P(\delta)$ (compare with their Fig. 5).
To proceed further, we have to compare the previous signature with
the one expected from a model of the LSS. As in \cite{Cuoco:2005yd},
we use the IRAS PSCz galaxy catalogue~\cite{saunders00a}. We address
to our previous work \cite{Cuoco:2005yd} as well as to the original
paper \cite{saunders00a} for technical details about the catalogue
and about the calculation of the UHECR sky map---which takes into
account energy losses as well---that we use in the following. It is
important reminding that the catalogue suffers of an incomplete sky
coverage. This includes a zone centered on the galactic plane and
caused by the galactic extinction and a few, narrow stripes which
were not observed with enough sensitivity by the IRAS satellite.
These regions are excluded from our analysis with the use of the
binary mask available with the PSCz catalogue itself. This reduces
the available sample (by about 10\%) to 93 events and the nominal
chance probability to 0.1\% (Fig. \ref{w_delta_check},
dashed--purple line). Note that this is a quality factor of the
catalogue, not an intrinsic problem of the data or theoretical
prediction. The green/dot--dashed line in Fig. \ref{w_delta_check}
shows the chance probability of the signature found in
\cite{Kachelriess:2005uf}, if the random events are sampled
according to the LSS distribution (obviously convolved with the
experimental exposures), rather than from an uniform one. Finally,
the dotted line shows the same result if the random events are
sampled according to {\it the square} of the LSS distribution, as
one would expect e.g. for a strongly biased population of sources.

\begin{figure}[!thb]
\vspace{-0.9pc}
\begin{center}
\begin{tabular}{c}
\epsfig{figure=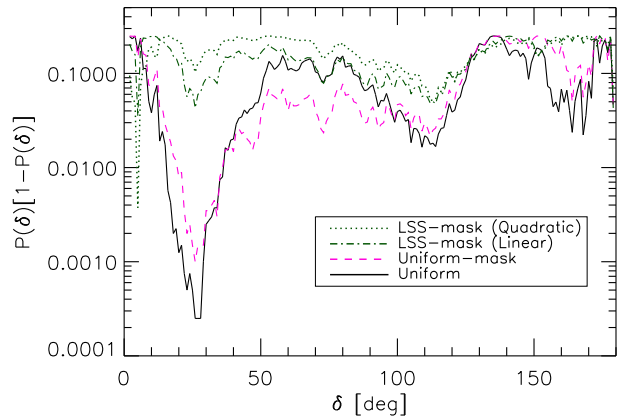,width=1.0\columnwidth}
\end{tabular}
\end{center}
\vspace{-0.9pc} \caption{The function $P(\delta)\times
[1-P(\delta)]$ for the same cases shown in Fig. \ref{w_delta_check}.
See text for details.} \label{fig2}
\end{figure}

The prominent minimum of \cite{Kachelriess:2005uf} is greatly
reduced when using as null hypothesis the LSS model instead of the
uniform one; this effect is even more prominent in the quadratic
map. Also, the data are less clustered than expected from an uniform
distribution at $\delta\sim 160^\circ$, where $P\sim 1$. This
additional puzzling feature disappears when using the LSS null
hypothesis, as it appears clearly in Fig. \ref{fig2}, where we plot
the function $P(\delta)\times [1-P(\delta)]$ for the same cases of
Fig. \ref{w_delta_check}. This function vanishes if any of $P$ or
$1-P$ vanishes and has the theorethical maximum value of $1/4$.
Thus, the higher its value is the more consistent the data are with
the underlying hypothesis. Apart from the very small scales, where
our results are unrealistic since we did not include magnetic
smearing or detector angular resolution, the better concordance of
the UHECR distribution with the LSS distribution than with the
uniform one is evident at any scale. Taken at face value, our result
implies a nominal probability $P\agt 5\%$ that the main signature
found in Ref. \cite{Kachelriess:2005uf} arises as a chance
fluctuation from the LSS distribution. This suggests that the
clustering properties of LSS are in much better agreement with the
experimental data than a pure isotropic distribution. This is not an
unexpected feature given that, as found in \cite{Cuoco:2005yd}, the
typical size on the sky of the clusters of structures lie in the
range 15$^\circ$-30$^\circ$.

The absolute scale of the curves shown in Fig. \ref{w_delta_check}
is affected by an uncertainty due to the true energy scale: we
calculated the map assuming that the HiRes energy scale is the
correct one, in agreement with Berezinsky et al.'s fit of the dip
due to pair production of protons on CMB \cite{Berezinsky:2005cq}.
But if the true energy is higher, as a compromise solution with the
other experiments may require, the chance probability is slightly
higher. In this respect, one may look at our result as a
conservative one. Hence, the largest sample of
\cite{Kachelriess:2005uf} which we chose on the basis of the
strongest signal is consistent within ``$2 \sigma$" with the
clustering properties expected from LSS
distribution\footnote{Consistent within ``$2 \sigma$" means here at
least in 5\% of the cases. The distribution is indeed far from
gaussian, and the number of $\sigma$ can be used in its loose sense
only.}. Also given the fact that the ``true probability" is higher
than the nominal one (due to the penalty factor of the search a
posteriori performed in \cite{Kachelriess:2005uf}), this may be
considered as an argument in favor of their interpretation.

\begin{figure}[!thb]
\vspace{-0.9pc}
\begin{center}
\begin{tabular}{c}
\epsfig{figure=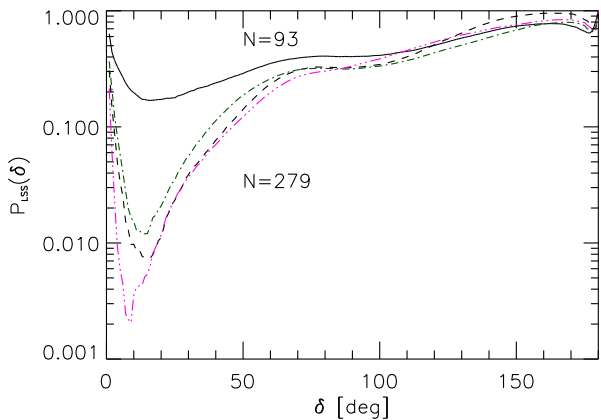,width=1.0\columnwidth}
\end{tabular}
\end{center}
\vspace{-0.9pc} \caption{Chance probability $P_{\rm LSS}(\delta)$ to
observe a larger value of the autocorrelation function as function
of the angular scale $\delta$ for two samples of 93 and 279 data
according to the LSS map cutted at $4\times 10^{19}\,$eV. The three
lower curves for the $N=279$ case show the signature when the map
cutted at $E>2,4, 5\times 10^{19}\,$eV---from top to bottom---is
used, while the upper solid curve refers to the case $N=93,\:
E>4\times 10^{19}\,$eV.} \label{w_delta_LSS}
\end{figure}

In order to understand why their result is consistent with our
sampled catalogue, it is useful to look at the chance probability of
the autocorrelation signal of events sampled from the LSS according
to the experimental exposures, assuming linear correlation. We show
this function in Fig. \ref{w_delta_LSS} for two samples of $N=$93
and $N=$279 data, respectively the same statistics of the dashed and
dot--dashed curves in Fig. \ref{w_delta_check}, and a factor 3
higher. The curves are obtained as follows: a large number $M(\simeq
10^4)$ Monte Carlo realization of $N$ events is sampled according to
the LSS probability distribution, and for each realization $i$ we
calculate the function $w_i^{\rm LSS}(\delta)$. We generate
analogously $M$ random datasets from an uniform distribution, and
calculate $w_j^{\rm iso}(\delta)$. We have thus $M^2$ independent
couples of functions $(i,j)$. The fraction of the $M^2$ simulations
where the condition $w_j^{\rm iso}(\delta)\geq w_i^{\rm
LSS}(\delta)$ is fulfilled is the probability
\begin{equation}
P_{\rm
LSS}(\delta)=\frac{1}{M^2}\sum_{i=1}^M\sum_{j=1}^M\Theta[w_j^{\rm
iso}(\delta)-w_i^{\rm LSS}(\delta)],
\end{equation}
which is the function shown in Fig. \ref{w_delta_LSS}.

An important qualitative feature is that the shape of the curve
presents indeed a broad minimum at scales of $\delta\alt 30^\circ$,
and a moderate plateau at scales of $70^\circ\alt \delta\alt
130^\circ$. As shown by the $N=279$ case, in particular the first
feature is intrinsic to the data: the more data are sampled, the
more enhanced it appears. This is also the trend shown in
\cite{Kachelriess:2005uf} when enlarging the experimental statistics
considered. Also, the higher the energy cut in the map, the stronger
the signature, since the local structures are more and more
prominent. Finally, the LSS data samples are typically less
clustered than the uniform ones at $\delta\agt 150^\circ$ ($P_{\rm
LSS}>0.5$).

On the other hand, the minimum found in Fig. \ref{w_delta_LSS} for
the sample of 93 data is much less prominent than the one shown by
the dashed curve in Fig. \ref{w_delta_check}. This explains why the
consistency is ``only'' at the level of $\sim 5\%$. This fact is not
unexpected, given the predictions of Ref. \cite{Cuoco:2005yd}: $\sim
100$ events are too few {\it to guarantee} a detection of the
imprint of the LSS with a high significance. Consistently with the
results of Fig. \ref{w_delta_check}, we checked that the dip in Fig.
\ref{w_delta_LSS} becomes more pronounced if we use a quadratic bias
with LSS. Although not statistically significant at the moment, a
confirmation of a highly clustered signal at intermediate scales may
suggest thus a more than linear correlation of UHECR sources with
the galaxy density field. Alternatively, the signal may be due to
the magnetic smearing of a few relatively strong point-sources.

Clearly, a smoking gun in favor of the LSS-distribution would be a
correlation between the data and the expected excess in the LSS map.
By performing an analysis similar to the previous one, but in terms
of the cross-correlation function between simulated data and sampled
ones, we did not find any evidence favoring a LSS origin with
respect to the uniform case. Actually this is not unexpected within
the model considered in \cite{Cuoco:2005yd}, since $\sim 100$ data
at energy $\agt 4\times 10^{19}\,$eV is still a too low statistics
to draw a firm conclusion in this sense. However, the lack of this
signature may also be related to the role of intervening magnetic
fields. Acting on an energetically (and possibly chemically)
inhomogeneous sample, magnetic fields may displace the observed
positions with respect to the original ones in a non trivial way,
without evidence for a characteristic scale, at least in a poor
statistics regime. A possible hint towards a non-negligible role of
magnetic fields is given also by the fact that the dip in the LSS
signal is already present at relatively small angles. This feature
may have disappeared in the UHECR sample due to a smearing effect of
the magnetic fields.

\section{Discussion and Conclusions}\label{conclusions}
We have analyzed the hypothesis that the broad maximum of the
two-point autocorrelation function of the UHECRs arrival directions
around 25$^\circ$ found in Ref. \cite{Kachelriess:2005uf} may be due
to the imprint of the LSS. We have concluded that this suggestion is
at least partially supported by the UHECR sky map constructed
starting from a LSS catalogue. Even their nominal (non penalty
factor-corrected) result for the autocorrelation function is
consistent within 2 $\sigma$ with our expectations. A stronger
correlation with source luminosity or a more-than-linear bias with
overdensity may improve the agreement. Also, the correlation may not
be directly with the LSS themselves: any class of sources which is
numerous enough is expected to show some indication in favor of this
correlation. The low statistics and the role of the magnetic field
deflections may explain why no significant {\it cross-correlation}
between data and LSS overdensities is found.

The authors of Ref. \cite{Kachelriess:2005uf} also claim that the if
the signal found is real, a heavy composition of the UHECRs is
disfavored. However, we note that a heavy or mixed composition of
the UHECRs may well be consistent with the signature. If we limit to
the role of the (relatively well known) galactic magnetic field, a
naive extrapolation of the simulations performed in
\cite{Kachelriess:2005qm} would indicate in the linear regime
deflections for iron nuclei of about $130^\circ$ with respect to the
incoming direction. UHE iron nuclei would then be in a transition
from diffusive to ballistic regime. Nonetheless, the signal is
sensitive to the {\it relative} deflections of ``bunches" of cosmic
rays originating from a similar region of the extragalactic sky, for
which typical models of the regular galactic magnetic field predict
a smearing $\alt 40^\circ$ even for iron nuclei, as long as their
energies do not differ by more than about 30\%. One may even
speculate that the second dip at large angles (arising from
cross-correlation of different groups of overdensities) might
originate from primaries of different rigidity coming from the same
few sources, splitted apart by intervening fields.  Also, the
consideration in \cite{Kachelriess:2005uf} that accounting for the
medium scale structure in UHECRs may change the significance of
claims of small-scale clustering should be carefully examined. If
the picture emerging from Ref. \cite{Kachelriess:2005uf} and this
paper is consistent, {\it both} LSS and magnetic fields play a role
in shaping the signal, otherwise it is hard to explain the lack of
cross-correlation with known overdensities of LSS. ``Filaments and
voids" in the observed data do not match the position of filaments
and voids in the LSS. But if they are nonetheless connected, this
difference must be rigidity-dependent. Thus, a cluster of events
with high rigidity may well arise in a void of the presently known
UHECR filamentary structure, which may sit closer to an overdensity
of the LSS. Indeed, the clustered component of the AGASA data
favoring small-scale clustering does show a different energy
spectrum than the non-clustered component. This discussion
emphasizes that, unfortunately, it is virtually impossible to draw
strong conclusions at present, even assuming that the clustering at
intermediate scales is physical.

In conclusion, the analysis performed in this paper does not exclude
that the signal found in \cite{Kachelriess:2005uf} may be due to the
imprint of the LSS, an indeed gives some support in this sense.
Definitely, the larger statistics that the Auger Observatory is
going to collect in the next years is needed to tell us finally if
astronomy is possible with UHECRs or, equivalently, if we will be
ever able to look at the sky with new and ``ultra-energetic''
eyes.\\{}\\

{\bf Acknowledgments}\\ We thank M. Kachelrie{\ss}
for useful comments. P.S. acknowledges support by the US Department
of Energy and by NASA grant NAG5-10842. This work was also supported
by the PRIN04 "Fisica Astroparticellare" of Italian MIUR.

\vspace{-1pc}

\end{document}